\documentclass[11pt,twoside]{article}
\makeatletter
\newcommand{\specificthanks}[1]{\@fnsymbol{#1}}% Inserts a specific \thanks symbol
\makeatother

\usepackage{latexsym}
\usepackage{amssymb, amsbsy, amsmath, amsfonts, amssymb, amscd, color}
\usepackage{graphicx}
\usepackage{float}
\newcommand{\commentout}[1]{}

\newcommand {\Chi} {{\bf \raise 2pt \hbox{$\chi$}} }
\newcommand{\beq}{\begin{equation}}
\newcommand{\eeq}{\end{equation}}
\newcommand{\bea} {\begin{array}{rl}}
	\newcommand{\eea} {\end{array}}
\newcommand{\bepa}{\left\{ \begin{array}{l}}
	\newcommand{\eepa} {\end{array}\right.}

\newcommand{\qed}{{ \hfill
		{\unskip\kern 6pt\penalty 500 \raise -2pt\hbox{\vrule\vbox to 6pt{\hrule width 6pt
					\vfill\hrule}\vrule} \par}   }}

\title{Employee turnover prediction and retention policies design: a case study.}
	\author{\scriptsize{Edouard Ribes}\thanks{Ecole Polytechnique,91128 Palaiseau, France,Email: edouard.augustin.ribes@gmail.com}\textsuperscript{ \specificthanks{5}}
	\and{\scriptsize{Karim Touahri}\thanks{Université 5 Descartes, Paris 13 Sorbonne City,France,Email: touahrikarim91@gmail.com}}
	\and{\scriptsize{Beno\^ \i t Perthame}\thanks{Sorbonne Universit\'es, UPMC Univ Paris 06, Laboratoire Jacques-Louis Lions  UMR CNRS 7598,  Inria, F75005 Paris, France}} 
	}

\begin{document}
\maketitle
\begin{abstract}
	This paper illustrates the similarities between the problems of customer churn and employee turnover. An example of employee turnover prediction model leveraging classical machine learning techniques is  developed. Model outputs are then discussed to design \& test employee retention policies. This type of retention discussion is, to our knowledge, innovative and constitutes the main value of this paper.
\end{abstract}

\noindent{\makebox[1in]\hrulefill}\newline
2010 \textit{Mathematics Subject Classification.} % 35K55, 35B25, 76D27, 92C50.
	\section{Introduction}
	Machine learning algorithms are often showcased in customer churn study. Applications in fields such as telecommunication or product marketing (gaming, insurance etc..)(see \cite{Ngai20092592},\cite{Vafeiadis20151} for a recent review) are multiple. The implementation of these methods in Customer Relationship Management is becoming the new norm, as improving customer retention yields superior business results. We argue that this type of techniques can easily be applied to employee turnover. Note that the employee turnover can actually be subdivided in 3 buckets: involuntary turnover (induced by the company), voluntary turnover (employee resignation) and retirements. Retirement and an involuntary turnover are out of the scope of this paper. Retirement is indeed generally legally enforced through specific local schemes and does not require a prediction. As for firing decisions, they are a company decision aiming at achieving the right workforce sizing or productivity levels. Therefore they do not require any anticipation on workers behavior. On the other hand, voluntary turnover is employee dependent. Mitigating it can be of interest to companies. This is even, in our opinion, one of the key workforce dynamic to understand in order to manage a company in a sustainable fashion (see \cite{Doumic2017} for example).\\
	Assume that a company $Y$ has a portion of its workforce in job $j$ that is deemed critical to achieve business results. Assume the job $j$ participates in a critical part of a product manufacturing process and is extremely difficult to source because of the required expertise and associated training time. Delays and/or failure in the manufacturing process can be extremely costly for the company, so that retention efforts may have to be considered to mitigate production risks. \\
	Retention is part of the standard Human Resources [HR] activity which is structured around the "5 Bs", namely : "Buy" and "Borrow"(=hire), "Build"(=train), "Bind"(=retain), "Bounce"(=fire). Retention investments usually factor individual performance, potential to grow in a company and are conjugated with a leave risk assessment. Let's assume that performance and potential are already assessed in a satisfactory fashion but that the leave risk is currently assessed by the employee's manager based on personal intuition. This managerial assessment is in reality a classification problem. The manager will indeed flag each year its employees has having either a "High Risk" (the employee will leave the company this year) or a "Low Risk" (the employee will not leave the company this year) of departure. Assume that the company $Y$ does not trust the managerial assessments because historically, they poorly correlate to the observed voluntary turnover fluxes. The company would therefore be interested in developing a prediction routine to structure its retention efforts.\\	
	The use of machine learning techniques for turnover prediction purposes has risen lately (see\cite{MLexample1} and \cite{MLexample2} for recent examples). Yet, to our knowledge, nothing has been done to design and test retention policies. In this paper, we will therefore leverage HR domain knowledge to build a turnover predictor and test retention policies. This paper will first explain how features can be engineered to feed a machine learning algorithm. In a second part, several machine learning methods and training sample techniques will be discussed and analyzed in terms of performance. Finally a sensitivity analysis will be carried out to design and test specific retention policies.
	
	\hfill  Important Legal Remark. The findings and opinions expressed in this paper are those of the authors and do not reflect the positions from any company or institution. Finally, please bear in mind that for confidentiality reasons numbers have been disguised in a way that preserves the same conclusions as the actual case study.
	
	\section{Feature Engineering.}
	\subsection{Previous Work.}
	%---------------------------------------
	Turnover is not easy to predict because it results from a combination of elements. So far, no consensus has been reached in terms of key elements to use. For example, an early review of voluntary turnover studies \cite{10.2307/258331} has found that the strongest predictors for voluntary turnover were age, tenure, pay, overall job satisfaction, and employee’s perception of fairness. But other studies have also stressed the importance of job performance \cite{5f9128bfb1f044a38337f4bd86d6a8da}, job characteristics (role, seniority in role...) \cite{ref1111}, enhanced individual demographic characteristics (age/experience,  gender, ethnicity, education, marital status) (\cite{ref222}, \cite{Finkelstein1}, \cite{Holtom1},  \cite{Hippel1}, \cite{Perterson1}, \cite{Sacco1}), structural characteristics (i.e. team size and performance) and geographical factors (\cite{RePEc:irs:iriswp:2006-04} , \cite{ref77777}).Finally some studies have also set an emphasis on salary, working conditions, job satisfaction, supervision, advancement, recognition, growth potential, burnout etc. \cite{Allen1}, \cite{Liu1},\cite{Swider1}, \cite{Heckert1}. And our list is far from being exhaustive...\\
	To our understanding, the key to successfully engineer features in machine learning exercises revolves around flexibility. Companies Human Resource Information Systems [HRIS] are famous for data inconsistency and quality issues. \textcolor{blue}{The absence of data standards thus calls for company specific engineering tasks.}
	\subsection{Turnover Domain Knowledege \& Data.}
	The dataset available for this study was discussed with a pool of peer-trusted experts among our clients group. The purpose of this discussion was to hand pick features that, in light of the previous bibliographical work, would be relevant for the given job $j$ and in the current historical context. Leveraging peer trusted experts was important because it enabled us to infuse domain knowledge to the project while creating buy-in in our client group for this kind of methodology. This was also critical as it helped us remove features that were considered non reliable. The selected features were:
	\begin{itemize}
		\item Localization. Individuals from a given country were selected. \textcolor{blue}{This country was described by a categorical feature representing the regional area where the employee was working. This was used to account for local labor market specific dynamics.}
		\item Knowledge. In order to account for specific domain expertise among job category $j$, the type of business unit in which the individual was working was added to the feature mix. 
		\item Individual data. Several features were suggested as far as individuals are concerned. First, time in the current position and tenure with a given company were added. Demographics in terms of age and gender were also highlighted. Additional elements specific to the dataset, such as performance \& potential evaluation, were discussed. Hierarchical level according the dataset grading structure was finally added to the mix.
		\item Individual managerial data.  Manager’s demographic information such as age and gender were added, as well as manager’s time in the current position and tenure in a given company. Manager's performance over the last year and average performance over the last 3 years \textcolor{blue}{were} also integrated.
		\item Individual team data. Team size and team percentage of high and low performers were added to the prepared mix.
	\end{itemize}
	The population under study in the given job category $j$ accounted for about 1000 employees in a given country. 2 years (referred to as year 1 and year 2) of data were available. The problem was of a classification nature as employees were either tagged as "Terminated" or "Active". Note that the dataset was curated of individuals who left involuntarily or because of retirement to respect the scope of the study.\\
	The full dataset was heavily unbalanced as the yearly number of people in job $j$ leaving voluntarily was about $20\%$.It was randomly split in a training and a test set, which were of equal size. \textcolor{blue}{Each of the two sets contained the same proportion of year 1 and year 2 information. This choice indeed yielded more robust performances than having the algorithms trained on one year and tested on the other.}
	\subsection{Feature Selection.}
	After this pre-processing, a selection layer was deployed in order to keep only the most relevant elements to the current problem out of the ones mentioned by the experts and the literature. Limiting the number of features used in machine learning problems is indeed crucial to properly manage the complexity of the learning phase and to avoid over fitting. In this paper, the mutual information (MI) criterion was used to assess the “information content” of each individual feature. This type of feature selection method is indeed well acknowledged and has proven useful for such topics (see \cite{298224} for an early example, \cite{Moldovan2016} for a customer churn example and \cite{Vergara2014} for a review.).
	\paragraph{Ranking with Mutual Information (MI).}
	Mutual Information (MI) is an essential metric of information and has been widely used for quantifying the mutual dependence of random variables (see \cite{MI1} for details).Formally the MI between an output variable $Y$ and an input variable $X$ is defined as:
	\begin{equation}
	MI(X;Y)=\sum_{x \in X}\sum_{y \in Y}P(x,y).\log{\dfrac{P(x,y)}{P(x).P(y)}}
	\end{equation}
	where $P(x)$ (resp. $P(y)$) is the probability associated to $X$ (resp. $Y$), and $P(x,y)$ is the joint probability of having both $X$ and $Y$.\\
	\paragraph{Features Filtering.}
	Several possibilities exist in order to select features. According to a recent survey \cite{Chandrashekar201416}, they can be summed up in three types of methods: filters, wrappers and embedded ones. Filter methods preprocess and rank features in order to keep only the best ones. In wrapper methods, the feature selection criterion is the performance of the predictor itself. The predictor is indeed wrapped on a search algorithm which will find a subset that gives the best predictor performance. Finally embedded methods include variable selection as part of the training process without splitting the data into training and testing sets.\\ 
	\begin{minipage}{0.5\textwidth}
		\begin{center}
			\includegraphics[width=\linewidth]{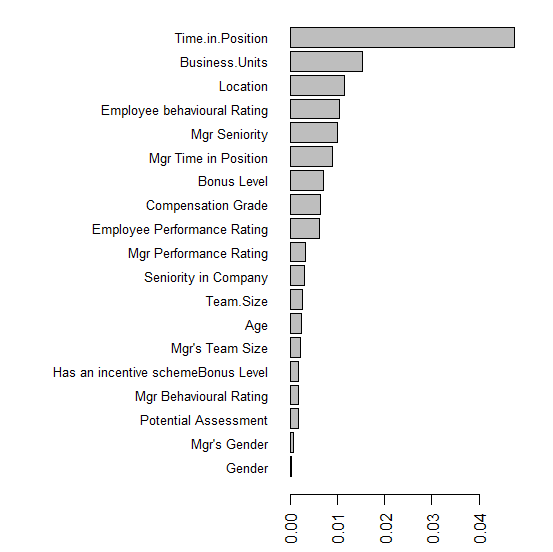}
			\title{Fig 1. Features Mutual Information}
			\label{fig:MIResults}
		\end{center}
	\end{minipage}
	\begin{minipage}{0.5\textwidth} In this case, a filter method was used based on the previously described mutual information criterion. This choice was motivated by its simplicity and the empirical efficiency of the approach. For each feature $X$, the mutual information $MI(X,Y)$ was compiled and only the top 60\% features were kept. Mutual information results are displayed below.
		Among the features of demographic nature, employee's gender and age as well as manager's gender were removed. Company bonus level information and structural hierarchical details proved useful but required granularity to be of use. Team size information disappeared. It is interesting to see that with the current dataset; only employee behavioral assessments relates to the turnover decision. Further investigation would be needed on this aspect of the dataset.
	\end{minipage}
	\\
	
	\section{Machine Learning Methods.}
	The analytical techniques deployed rely on well-known standards such as SVM, random forests or KNN that have been proposed to predict turnover (see \cite{Sikaroudi2015} or \cite{MLexample1} for examples). They have been adjusted to account for class imbalance and potential associated biases.
	\subsection{Class Imbalance Correction.}
	Class Imbalance is a common theme on customer churn prediction (\cite{DBLP:journals/corr/abs-1106-1813}, \cite{Amin2015}...), which is a topic similar to the one developed in this paper. Standard solutions involve subsampling techniques to smooth the disparities between the observed classes. In this paper, several methods were tested:
	\begin{itemize}
		\item Down-sampling: sample the majority class to make its frequency closer to the rarest class. 
		\item Up-sampling: resample the minority class to increase its frequency.
		\item Weighting: place a heavier penalty on misclassifying the minority class \cite{weights} by assigning a weight to each class, with the minority class being given a larger weight (i.e., higher misclassification cost). 
		\item "SMOTE": over-sample the minority (rarest) class and under-sample the majority class (\cite{SMOTE}).
		\item "ROSE": generate new artificial data from the classes to correct class imbalance according to a smoothed bootstrap approach \cite{Menardi2014}.
	\end{itemize}
	\subsection{Selected Classification Methods.}
	The classification problem was approached with the following techniques:
	\begin{itemize}
		\item Naive Bayes (NB) classifiers (see \cite{xia2008model} for an application) calculate the probability that a given input sample belongs to a certain class. Assuming  a number of classes $Y = (y1, y2, … , yk)$ and a given a unknown sample X, a Naive Bayes classifier assigns the sample X to the class with the highest probability $C=ArgMax_{y\in (y1...yk)}(p(y|X))$		
		\item Linear Discriminant Analysis (LDA) is a generalization of Fisher's linear discriminant analysis. It can be explained as computing a z-score, which is then used to estimate the probabilities that a particular member or observation belongs to a class.
		\item Support vector machines (SVMs) achieve class separation by finding (in their linear version) a hyper plane in a high dimensional space. The intuition is that a good separation is achieved by the hyper plane that has the largest distance to the nearest training data points of any class. The larger the margin is, the lower the generalization error of the classifier. For this reason, SVMs are also referred to as maximum margin classifiers. Multiple versions exists (linear, polynomials, gaussian...) (see  \cite{Gent06churnprediction})
		\item Random Forests (RFs) \cite{RandomForest} result from the combination of tree classifiers. Each tree depends in the values of a random vector sampled independently and with the same distribution for all trees in the forest. The generalization error of a forest of tree classifiers depends on the strength of the individual trees in the forest and the correlation between them. RF has empirically achieved very good classification performance in numerous cases and is robust against over-fitting.
	\end{itemize}
	Note that the R Caret package \cite{RCaret} was the library used to support the study.
	\subsection{Evaluation \& Results.}
	\paragraph{Cross-Validation.}The dataset was split 80:20 into training and hold out sets in a 10-fold cross validation scheme. For each algorithm, a grid-search was performed over the tuning parameters, including regularization or penalty hyper-parameters. 
	\paragraph{Performance Metric.}The Area under the receiver operating characteristic curve (ROC-AUC) was the chosen performance measure. The AUC is a general measure of accuracy. It decouples classifier assessment from operating conditions i.e., class distributions and misclassification costs \cite{Lessmann1}. It is better suited to non-balanced classification problems than standard accuracy or error test as it allows a graphical representation of specificity (false positive rate) and selectivity (true positive rate) of the classifiers on the dataset. Note that, in this case, the specificity can be interpreted as the percentage of workers flagged as leavers that really left, while the selectivity represents the percentage of workers appropriately flagged that stayed in their company.
	\paragraph{Results.}We obtained the following results:\\
	\begin{minipage}{0.5\textwidth}
		In terms of imbalance correction, we found that up sampling and ROSE sampling yielded the best performance in this case. As expected, when imbalance was not corrected, algorithms tended to favor the majority class to achieve the best performances.
	\end{minipage}
	\begin{minipage}{0.5\textwidth}
		\begin{center}
			\title{Table 1. Classification Performance among resampling methods.}
			\begin{tabular}{c c c c c} % centered columns (4 columns)
				& & & & \\
				\hline\hline %inserts double horizontal lines
				Sampling & Best Algorithm & ROC & Spe. & Sel. \\
				\hline
				None & LDA & 0.73 & 0.94 & 0.19 \\
				Down & LDA & 0.74 & 0.61 & 0.67\\
				Up & RF & 0.96 & 0.82 & 0.96\\
				Rose & RF & 0.95 & 0.81 & 0.96\\
				Smote & RF & 0.87 & 0.86 & 0.74\\
				\hline
			\end{tabular}
		\end{center}
	\end{minipage}
	\\
	\begin{minipage}{0.5\textwidth}
		\begin{center}
			\includegraphics[width=\linewidth]{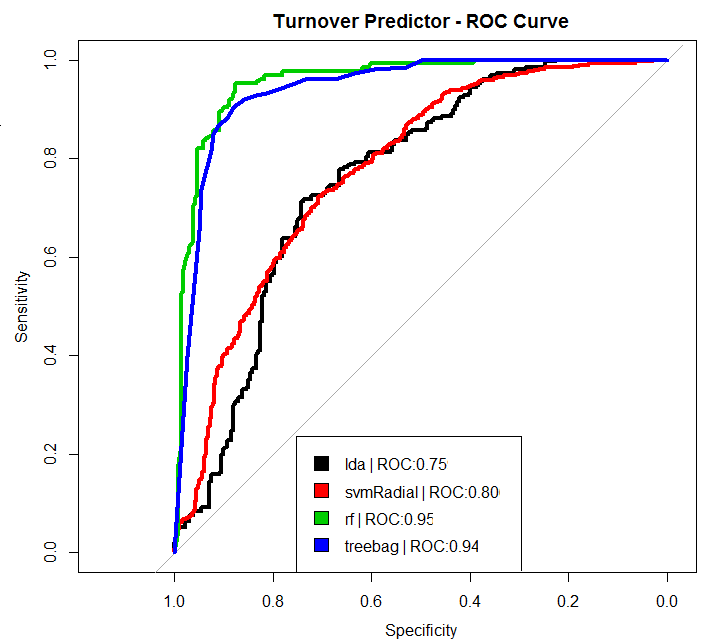}
			\title{Fig 2. ROC Curves with ROSE resampling}
			\label{fig:ROCCurve}
		\end{center}
	\end{minipage}
	\begin{minipage}{0.5\textwidth}
		In terms of algorithms, we saw that the tree based ones performed best. The following performances were indeed recorded with ROSE resampling:
		\begin{center}
			\title{Table 2. Classification Performance among algorithms.}
			\begin{tabular}{c c c c} % centered columns (4 columns)
				\hline\hline %inserts double horizontal lines
				Algorithm & ROC & Specificity & Selectivity \\ [0.5ex]
				\hline
				LDA & 0.75 & 0.67 & 0.74 \\
				SVM (Radial) & 0.80 & 0.69 & 0.76 \\
				RF & 0.95 & 0.81 & 0.96 \\
				Treebag & 0.94 & 0.78 & 0.94 \\
				\hline
			\end{tabular}
			\bigskip
		\end{center}
		The corresponding ROC Curves have been reported to the left.
	\end{minipage}
	\\
	\begin{minipage}{0.5\textwidth}
		As far as the variables of importance are concerned, we saw that the employee behavior and performance (especially when rated low) are important elements to the turnover behavior prediction. Structural elements in terms of compensation schemes and business units’ specificities were also rated high. The same goes for certain seniority in company and time in position year bands.\\
		According to the graph \ref{ fig:VarImp }, new joiners in the company are at risk. So are people that have been in their roles for more than 4 years. This last element may relate to a perceived lack of career opportunity. This seems to be corroborated by the importance of managerial time in position. Interestingly managers’ importance in turnover decisions appears to be mainly related to their seniority in the company or their time on the job. Manager performance or behavior does not rate extremely high.\\
		Another interesting element is that remote jobs do not seem to be driving more turnover than other offices location. This would hint to local labor market specificities.
	\end{minipage}
	\begin{minipage}{0.5\textwidth}
		\begin{center}
			\includegraphics[width=\linewidth]{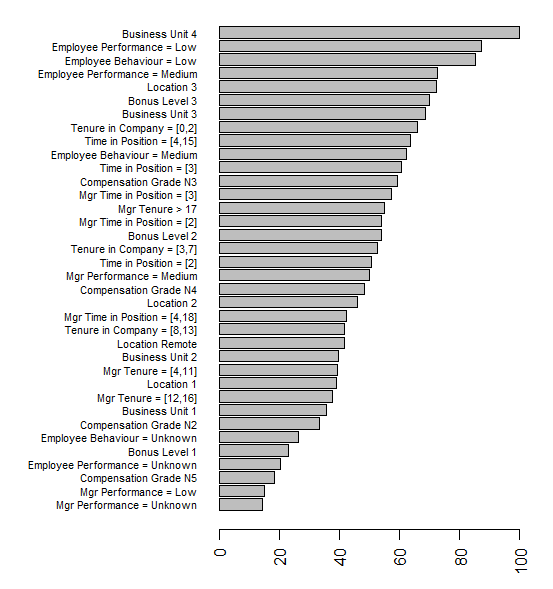}
			\title{Fig 3. Varibale Importance (RF with Rose Resampling)}
			\label{fig:VarImp}
		\end{center}
	\end{minipage}
	\\
	\\
	In this section, it has been showed that it possible to model voluntary turnover with a couple of standard variables. Even though this may not hold for all types of jobs, the announced performance proved replicable on the \textcolor{blue}{prepared} test set. A discussion about the model use to deliver insights on retention policies is now in order.	
	\section{Retention Discussion.}
	According to recent reviews in the field of employee retention (see \cite{das2013employee}, \cite{ramlall2004review}), the results from the previous section are not surprising. A number of generic actions to bind people to firms have already been discussed over the years. Compensation topics, including rewards and recognition schemes are among the most popular (see \cite{bryant2013compensation} for an example). Other elements such as opportunity for growth \& development, accountability and leadership structure have also been stressed. Some further elements regarding the workplace and work life balance have been raised too. \\But as stressed by the literature reviews retention efforts should be tied to an in depth turnover analysis, \textcolor{blue}{which will be the aim of this final part of our study}. The previous results motivated us to test five simple possible programs. The first two are about reviewing the job locations: in the program $P1$, remote jobs are reassigned to the location 1, while in $P2$ jobs in location 3 are reassigned to location 1. The next ones are about the managers for the considered labor categories. In the program $P3$, all managers were forced to have a first internal experience prior to their managerial role. This translated in reallocating all managers with company tenure between 0 and 2 years to a 3 to the 7 years band. In the program $P4$, managers were assumed to rotate frequently between team so that their time in position would be kept below 2 years. Finally we tested an employee program $P5$ that assume that people were bound to the company during their first 2 years in position.
	\subsection{Retention Policy Simulations.}
	The five programs mentioned ($P1$ to $P5$) were tested on a prediction set, which was built from the current year of data where no turnover indication was available. It was therefore different from both the test and training set mentioned earlier. Size and characteristics of the prediction set were similar. Certain parameters of the prediction set were altered to perform the sensitivity analysis required to simulate the retention programs as previously described. Two types of policies were investigated. It is indeed either possible to apply the program $P1$ to $P5$ to all the people in scope of the study or to target only the individuals flagged as leavers and adopt a policy that suits them best.
	\paragraph{Mass Actions.} As far as mass actions are concerned, the following table summarizes our findings:
	\begin{table}[!t]
		\caption{Retention Mass Actions Simulations.}
		%\label{table_example}
		\centering
		\begin{tabular}{c c c} % centered columns (4 columns)
			& &    \\
			\hline\hline %inserts double horizontal lines
			Program Type & Program Details & $\%$ of leavers \\
			\hline
			\hline
			None & No Retention Policy & 41.5$\%$ \\
			\hline
			P1.Location & Remote Job reassigned to Location 1
			& 40.9$\%$ \\
			P2.Location & Location 3 reassigned to Location 1 & 41.3$\%$ \\
			\hline
			P3.Manager & People Managers had 1 role internally
			& 39.7$\%$\\
			P4.Manager & Manager Rotation Program & 32.3$\%$\\
			\hline
			P5.Employee & Bind newly assigned people
			& 30$\%$ \\
			\hline
		\end{tabular}
	\end{table}
	
	The results displayed above show that programs related to internal mobility are among the most efficient options to reduce turnover. If the program $P5$ appears the most efficient, it also is the most difficult to set in place and offers little upside compared to the managerial rotation program. Interestingly location change offers little value from a retention standpoint. This could be explained in this case by the fact that the labor category at stake can work in a remote fashion. This indeed creates a country wide labor market to compete in.
	\paragraph{Targeted Actions.} According to the algorithm results, about 40 \% of the population is likely to leave and requires an action. The investigated course is now to test for each of the supposed leavers which of the programs($P1$ to $P5$) is useful. This will limit the amount of effort required while optimizing the employee retention. This leads to a final \% of leavers of 24\%, which represents a significant mitigation of turnover. The results in terms of actions are summarized in the table below:
	\begin{table}[!t]
		\caption{Retention Targeted Actions Summary.}
		\centering
		\begin{tabular}{c c c} % centered columns (4 columns)
			& &  \\
			\hline\hline %inserts double horizontal lines
			Action Type & $\%$ of the total  & $\%$ of the potential  \\
			&Population concerned & leavers concerned\\
			\hline
			\hline
			None & 82,42\% & 57,67\%  \\
			\hline
			P1.Location & 0,88\%	& 2,12\%\\
			P2.Location & 0\% & 0\% \\
			\hline
			P3.Manager & 	1,54\%
			& 	3,70\%\\
			P4.Manager & 10,55\% & 	25,40\%\\
			\hline
			P5.Employee & 4,62\% & 11,11\%\\
			\hline
		\end{tabular}
	\end{table}
	These results show that to halve the turnover rate, actions only need to be started for ~18\% of the total population. In the end, about 25\% of the population at risk of departure needs a new manager and 11\% require an incentive to stay over the first two years of their role. This highlights solutions that the Human Resource organization and notably the talent management centers of expertise could deploy in in order to successfully \textcolor{blue}{contribute to one’s organization management.}
	\subsection{Discussion.}
	Several limitations to this study exist and are listed below:
	\begin{itemize}
		\item \textbf{Model Limits.} First, we did not have a holistic vision on individual parameters. For instance, \textcolor{blue}{detailed} compensation parameters were missing. It is completely possible that fed with another set of features, the algorithms would have yielded better performances and different interpretations, for instance in terms of importance ranking. The story developed in this study case results from insights driven discussion between experts. We argue that if advanced analytics are important, the key elements here were the discussions that resulted in a general buy in of the retention policies simulations.
		\item \textbf{Emotional Component in the job relationship.} The discussions held with our sounding board raised another issue. Turnover may stem from a mix of rational and emotional components. The dataset was mainly organized around standard organizational or labor indicators. Little to no indication regarding people profiles, behaviors or preferences was available.
		\item \textbf{Stress Supply vs Demand Approach.}
		The developed approach works well to propose turnover mitigation measures. However we believe that it should not be the first thing to do. Instead we recommend leading the turnover and retention discussion with thorough job level productivity discussion. This indeed helps companies navigate in the supply \& demand type of environment that are labor markets and therefore helps them define what an acceptable turnover rate is. Productivity will indeed help defining the cost of the turnover related disruption while the supply demand equilibrium will help finding the right solution mix in terms of staffing vs retention efforts. We personally believe that the human resource function is a key operating body of companies. Yet it needs to tie its activity and services to tangible business outcomes (sales, cost, productivity) to prove its value. This wasn't stress enough in our study and a review of productivity definition across functions appears mandatory to standardize this type of approach.
	\end{itemize}
	
	%------------------------------------------------
	\section{Conclusion \& Next Steps.}
	In this paper, a method to approach employee retention has been proposed using standard machine learning techniques. In a first part, a \textcolor{blue}{literature} review has shown how close this problem is from the customer churn one. Yet, to our knowledge, little has been done regarding employee turnover from a quantitative standpoint. This generated a discussion with experts on how to tackle this problem and generate\textcolor{blue}{meaningful} features. Those features were then fed to standard machine learning algorithms under various resampling schemes to account for classes’ imbalance. The best performing algorithm was retained to predict turnover and a sensitivity analysis was led to understand the effects of several retention policies. We found that for the data at hand, the most efficient retention practice was to develop talent mobility across positions.\\
	The principal next step to this study would be to include \textcolor{blue}{detailed} compensation elements to the feature mix. The current model is indeed more around the employee perception of its internal environment than about rational economic tradeoffs. We indeed believe that such models have the potential to mimic an individual job related utility function. 
\bibliographystyle{ieeetr}
\bibliography{bibli}
\end{document}